\documentclass[12pt]{article}

\usepackage[utf8]{inputenc}
\usepackage[T1]{fontenc}

\usepackage{lmodern}

\usepackage{amssymb,amsmath}

\usepackage{bm}
\usepackage{bbm}

\newcommand{\blind}{0}

\usepackage{url}
\usepackage{float}
\usepackage{booktabs}
\usepackage{graphicx}

\usepackage{setspace}
\usepackage{natbib}

\usepackage{authblk,color}

\usepackage{longtable}
\usepackage{multirow}

\usepackage{hyperref}

\newcommand{\bbeta}{{\mbox{\boldmath $\beta$}}}

\newcommand{\y}{{\mbox{\boldmath $y$}}}

\newcommand{\bepsilon}{{\mbox{\boldmath $\epsilon$}}}

\newcommand{\bi}{\begin{itemize}}
\newcommand{\ei}{\end{itemize}}

\newcommand{\zero}{{\mbox{\boldmath $0$}}}

\date{}
\begin{document}

\def\spacingset#1{\renewcommand{\baselinestretch}%
{#1}\small\normalsize} \spacingset{1}

%%%%%%%%%%%%%%%%%%%%%%%%%%%%%%%%%%%%%%%%%%%%%%%%%%%%%%%%%%%%%%%%%%%%%%%%%%%%%%

\if0\blind
{
  \title{\bf Anomaly Detection in Energy Usage Patterns}
\author[1]{Henry Linder}
\author[1,*]{Nalini Ravishanker}
\author[1]{Ming-Hui Chen}
\author[2]{David McIntosh}
\author[2]{Stanley Nolan}
\affil[1]{Department of Statistics, University of Connecticut, Storrs,
  CT 06269}
\affil[2]{Utility Operations \& Energy Management, Facilities Operations, University of Connecticut, Storrs,
  CT 06269}
\affil[*]{Corresponding author: \texttt{nalini.ravishanker@uconn.edu}}
  % \author{Henry Linder
  %   % \thanks{
  %   % The authors gratefully acknowledge \textit{please remember to list all relevant funding sources in the unblinded version}}
  % \hspace{.2cm}\\
  %   Department of Statistics, University of Connecticut\\
  %   and \\
  %   Author 2 \\
  %   Department of ZZZ, University of WWW}
  \maketitle
} \fi

\if1\blind
{
  \bigskip
  \bigskip
  \bigskip
  \begin{center}
    {\LARGE\bf Anomaly Detection in Energy Usage Patterns}
\end{center}
  \medskip
} \fi

\bigskip
\begin{abstract}
 Energy usage monitoring on higher education campuses is an important step for
providing satisfactory service, lowering costs and supporting the move to green energy.
We present a collaboration between the Department of Statistics and
Facilities Operations at an R1 research university to develop statistically based
approaches for monitoring monthly energy usage and proportional yearly
usage for several hundred utility accounts on campus. We compare the interpretability and
power of model-free and model-based methods for detection of
anomalous energy usage patterns in statistically similar
groups of accounts. Ongoing conversation between the academic and operations teams enhances the practical utility of
the project and enables implementation for the university. Our work highlights an application of thoughtful and continuing  collaborative analysis
using easy-to-understand statistical principles for real-world deployment.
%  We present a collaboration between the Department of Statistics and
%  Facilities Operations at an R1 research university. We develop two
%  appraoches approach to monitor monthly energy usage on campus in
%  several hundred utility accounts, both based on proportional yearly
%  usage at the account level. We compare the interpretability and
%  power of model-free and model-based methods for detection of
%  anomalous energy usage patterns, applied to statistically similar
%  groups of accounts. We give discussion of the practical value of
%  various approaches, and provide details of implementation for the
%  university. Our work highlights the application of careful analysis
%  using simple principles for real-world deployment, rather than more
%  complex approaches that are also less accessible.
\end{abstract}

\noindent%
\vfill

\newpage
\spacingset{1.45} % DON'T change the spacing!

\section{Introduction} \label{sec:intro}

Energy management at large academic institutions usually poses unique
operational challenges \citep{cruz2017analyzing}. The
energy  infrastructure
%to support 
in a university can be  %diverse 
complex across multiple dimensions:
the number of buildings at a campus is large, the facilities exhibit
diverse use-types, and the buildings vary in size, energy efficiency,
and state of repair.
For utility managers, close and in-depth
scrutiny of energy usage across the entire campus must be balanced
against time and manpower
by using resources for the most critical maintenance problems.
The  balance between these priorities is the focus of Facilities Operations  in an institution.
Understanding and monitoring patterns in historical energy usage data and identifying
anomalous behavior is an important step in this direction, and statistical methods can provide a systematic  framework
for implementing these.

%The balance between these priorities is 
This is the focus of an ongoing
collaboration at the University of Connecticut between the Department of Statistics
and Facilities Operations -     %This consists of 
for working together in a
concerted way to achieve systematic and statistically valid
procedures to understand patterns and detect anomalies in energy usage, ultimately minimizing wasteful
consumption.
We %consider 
describe the problem of managing and monitoring monthly energy
usage in a large set of utility accounts located across the
university's $4000$ acre main campus.  The number of accounts is
prohibitively large to permit manual assessment by Facilities Operations   %building
engineers. Moreover, while each account individually represents only a
small fraction of overall energy consumption on campus, nonetheless,
when aggregated across several hundred accounts over the full fiscal
year, abnormal consumption may pose a substantial financial cost for
the university.  Since energy use is a fundamental need for university
operations, the mandate from Facilities Operations is to statistically
distinguish abnormal and unnecessary usage from anticipated and
essential usage. We do an in-depth analysis of time series of \textit{monthly} energy usage, as well as
monthly proportions within each year of the energy usage. We
have chosen to employ simple, but effective  %rather than complex 
statistical
procedures   %These methods 
that can be easily communicated to engineers and be jointly
implemented with our collaborators from Facilities Operations on ready-to-use dashboards. The variety of
building use-types on the university campus poses a unique challenge.
%in this context.  
An individual building's energy consumption profile
depends directly on factors including %building 
use-type, square
footage and building age.
%Despite this, year-to-year variation exhibits  
There also exists significant correlations between energy usage on different  campus buildings  due to common
drivers such as student and faculty population sizes,  climate
variations within a year, and campus-wide energy initiatives.

%Previous work 
Existing literature on methods for visualizing campus energy usage and
understanding and modeling energy data varies in
the level of detail and sophistication.
%of the methods. 
Students at the Worcester Polytechnic Institute \citep{ohara2007}   %outlined and 
discussed the physical infrastructure needs for real-time
collection of electricity data, with a focus on the hardware metering
technology used to measure demand.  The Harvard Medical School implemented
a real-time visualization dashboard for energy usage by building
\citep{lieberman2010}   %The dashboard presented 
to present graphics of historic
energy usage, with a goal of increasing community awareness.
\cite{breyer2011assessing} considered software solutions to encourage
behavioral changes in campus energy consumers at the University of
Michigan.  They focused on public-facing tools to increase awareness
among student, faculty, and staff communities to achieve structural
changes in demand for energy. \cite{ma2015energy} gave a descriptive analysis of aggregate
energy consumption at seven universities across the world.  They
considered the effects of population and building footprint size on
energy consumption, and used this to compare an energy consumption
index between universities to assess their relative energy usage.

%%%%%%%%%%%%

To our knowledge, existing literature on anomaly detection methods for energy consumption
seem to primarily  focus on high-frequency demand/usage data which are not suitable for
coarser-grained monthly data.
The methods used in these papers ranged from using  $z$-scores relating the mean and standard deviation for individual
time series based on previous behavior \citep{seem2007using} to looking for or large discontinuities in usage \citep{zhao2014novel} to using reduced dimension processes and looking for large prediction errors \citep{ma2017real} or distance based abnormality scores \citep{rashid2018monitor}.

The literature does not address statistical concepts such as power of
the anomaly detection schemes or the probability of false positives,
nor the mechanisms for anomaly detection in monthly proportions data.
It also does not directly address the managerial problem of monitoring
a large number of energy accounts, observed at a low frequency and
with small sample sizes.  Many existing methods are intended for
monitoring a small number of high-frequency time series, but that type
of data is only available for a small subset of buildings.  Moreover,
the Facilities Operations management problem exists even when high-frequency
meter data is unavailable, so small data approaches are still
necessary.

In this article, we describe our collaborative work with Facilities Operations which highlights the
value of statistical practice by academics
that focuses on operational problem-solving.
To identify candidate buildings that may exhibit aberrant energy
usage, we consider energy consumption across the fiscal
year.  We also integrate data from external sources to normalize usage
by relevant weather covariates, and then convert these into monthly usage
proportions within the year.  We group buildings according to similar
characteristics, and apply statistical clustering methods.  This
enables full utilization of the similarities in consumption profiles
across accounts.   % especially after the proportional transformation.
By grouping buildings with similar
energy consumption profiles, we obtain an aggregate benchmark against
which we assess the usage characteristics of individual accounts
within these buildings.
We propose a graphical,
model-free method for anomaly detection based on    %boxplots. 
%We construct 
approximate, group-wise control limits on %the basis of 
boxplots constructed for transformed monthly usage proportions data.  % on the basis of which we 
%We propose a graphical,
%model-free method for anomaly detection based on boxplots.
 We also propose an approach based on a linear model,
which we use as a benchmark against which to compare our graphical method.  Our model is motivated by a
statistical process control approach described by
\citet{fu2014spc}. They used historical
in-control data to estimate nuisance parameters in their model, leading to an
approximate Bartlett-type likelihood ratio (ABLR).
We report to Facilities Operations
engineers a list of accounts  flagged as potential
outliers. These anomalies can be investigated by Facilities Operations
on an ongoing basis for physical malfunctions or other maintenance
problems.a

The format of the article follows. We first provide
a description of a monthly gas consumption data set collected by
Facilities Operations in Section~\ref{sec:data-desc}.  In
Section~\ref{sec:process}, we analyze monthly proportional usage
within a fiscal year. in Section~\ref{sec:known-group}, we identify
anomalous accounts within known homogeneous groups, using two
approaches. First, we consider a model-free approach based on the
boxplot to identify energy accounts that are anomalous relative to
similar accounts. In Section~\ref{sec:model}, we introduce a
model-based approach to identify changes in mean value, and compare
its properties to those of the less sophisticated, model-free method..
In Section~\ref{sec:cluster}, we describe an extension to the setting
where the groups that partition accounts are not all known \emph{a
  priori} and must be statistically determined. Finally, we end in
Section~\ref{sec:implementation} with a discussion of the interface
that we provide to Facilities Operations.

\section{Data Description} \label{sec:data-desc}

% This section discusses four aspects about the data: removing invalid
% accounts; normalizing the consumption values; explain the nature of
% uneven length time series; missing value imputations.

The raw data consist of monthly measurements of natural gas
consumption collected by Utility Operations and Energy Management in
Facilities Operations at the university's main campus. A total of
$245$ separate utility accounts are available across 115 buildings,
with the number of accounts varying between buildings. In some cases
one building contains multiple accounts, such as an apartment building
with several units. In what follows, we have denoted buildings by
generic names due to privacy considerations. %to preserve privacy.

The availability of historical data varies by %depends on the 
building and
account, with the earliest observation in February 2007 and the last
in December 2018. Prior to the analysis, we omitted four accounts
which have missing final observations, as this is an indcation that
the accounts are no
longer actively used. We also omitted 1 account with %more than
over $10\%$ of observations missing, and 2 other accounts known
%beforehand 
\emph{a priori} to be substantial outliers. Table~\ref{tab:dist} shows the
time series lengths in our dataset, most of which are either 82 or 143
months long. For example, of the $71$ accounts in Apartment Complex A,
$69$ series were observed from February 2007, or for $143$ months; 1
series for $137$ months; and the last for only $82$ months.

\begin{table}[ht]
  \centering
  \begin{tabular}{llrllr}
  \toprule
Year & Month & Freq. & Year & Month & Freq. \\
  \midrule
2007 & February & 128 & 2011 & March &   3 \\
  2007 & May &   1 & 2011 & April &   2 \\
  2007 & August &   1 & 2011 & June &   1 \\
  2008 & May &   1 & 2011 & July &   2 \\
  2009 & April &   1 & 2012 & March &  83 \\
  2009 & December &   1 & 2013 & March &   1 \\
  2010 & March &   3 & 2013 & May &   1 \\
  2010 & April &   1 & 2015 & January &   3 \\
  2010 & September &   1 & 2015 & April &   2 \\
  2010 & October &   1 & 2016 & June &   1 \\
   \bottomrule
  \end{tabular}
  \caption{Series start dates with observed frequencies. Values
    count the series beginning in a given month.}
  \label{tab:dist}
\end{table}

\begin{figure}[H]
  \centering
  \includegraphics[width=\textwidth]{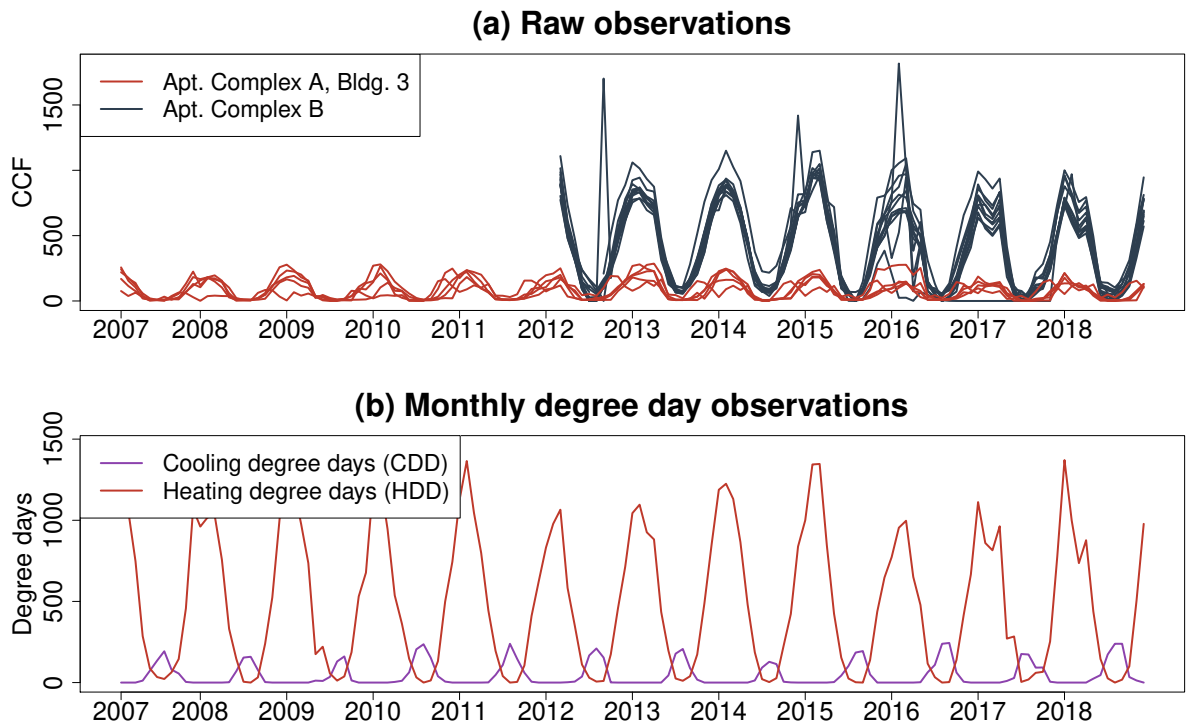}
  \caption{\emph{(a)} Raw data observed in Building C of
    Apartment Complex A (red lines), and all of Apartment Complex B
    (black). Data in Complex A is available since 2007, compared with
    2012 for Complex B. \emph{(b)} Historical degree day
    observations, by heating and cooling}
  \label{fig:dataraw}
\end{figure}

% Normalized observations
For the $238$ accounts used in the analysis, monthly gas consumption
was measured in units of hundred cubic feet (``CCF'').  Each
observation corresponds to a single utility bill in one account in one
billing period, measured as the difference between two meter
readings. The bills are grouped across accounts by a calendar-monthly
billing period.  The billing period provides a reference alignment of
each utility bill, despite differences across accounts in the specific
start and end dates in a given billing period. The monthly
observations for two distinct residential complexes are shown in
Figure~\ref{fig:dataraw}(a). Specifically, five accounts in Apartment
Complex A, Building C, and 12 accounts from Apartment Complex B are
shown.

It is possible that a value in any month for any account may be
exactly 0, for one of two reasons.  First, a building may be
unoccupied or otherwise inactive for that month. For instance,
Apartment Complex A, Building 6, has values of 0 in the
same months across several years, July and August. This is because the
residence hall is unoccupied in the summer. Second, the gas utility
company only charges for integer-valued CCF values, so that a recorded
value of 0 may actually represent a nonzero CCF value that is below 1.

In addition to energy usage, our analysis also includes weather
covariates. We used daily weather data from the National Oceanic and
Atmospheric Administration (NOAA). This data is made available through
the Applied Climate Information System
(\url{http://www.rcc-acis.org/}) web API. Based on guidance from
Facilities Operations, we obtained daily observations of heating
degree days (HDD) and cooling degree days (CDD) defined as
\begin{equation}
\text{HDD} = \max(0,65-\overline{u}), \mbox{ and } \mbox{CDD} = \max(0,\overline{u}-65),
\end{equation}
where $\overline{u}$ refers to the average (of the maximum and minimum
temperatures) on any given day.
Degree days provide a nonlinear measure of
temperature deviation from a neutral balance point (65 degrees), and provide a
robust context for temperature \citep{quayle1980heating}.  It is worth
noting that the daily value of a single degree day will generally be
greater than 1.  The degree day converts the one-day deviation
 from the
balance point $\overline{u}$, to $\overline{u}$ days of 1-degree
deviation from the balance point. Therefore, the degree day measures
the magnitude of a deviation from the balance point.
%Following the standard convention, we defined HDD and CDD as
To reckon the HDD (CDD) for a month, it is usual to sum the daily HDD
values. Figure~\ref{fig:dataraw}(b) shows the historical
observations of heating and cooling degree days.

In addition to the different number of calendar days in each month,
the specific start and end dates for energy consumption within any
month are not consistent across all accounts. We re-normalized the
energy data to account for this problem in Section~\ref{normalize}. We
also adjusted for weather effects.

\section{Processing Data and Grouping Accounts} \label{sec:process}

\subsection{Re-Normalization and Weather Adjustment of Data} \label{normalize}

\begin{figure}[H]
  \centering
  \includegraphics[width=\textwidth]{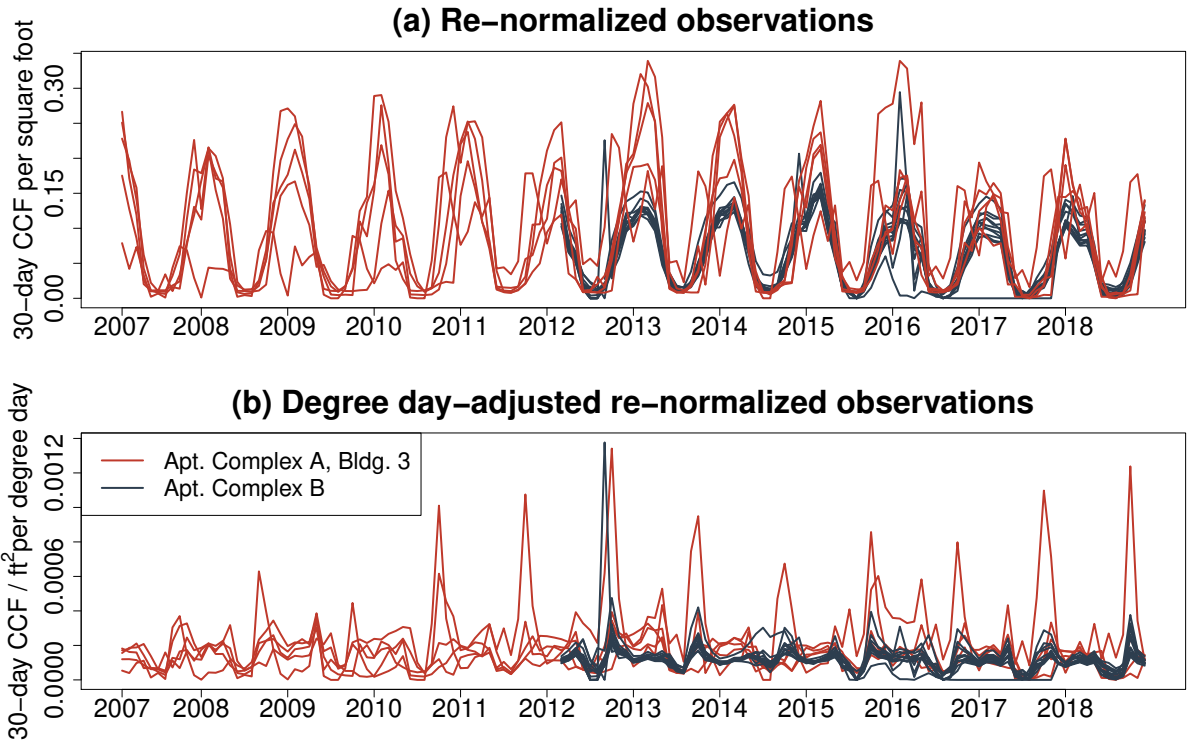}
  \caption{\emph{(a)} Normalized data, adjusted for nominal month
    billing period duration and per-unit square footage.  \emph{(b)}
    Degree day-adjusted re-normalized data, with imputed values.}
  \label{fig:datanorm}
\end{figure}

We calculated the average consumption per day for each observation,
and then re-normalized all observations to represent a 30-day month.
Furthermore, we applied a standard adjustment by dividing the CCF
value by the area of the building (square footage), so that each
observation represents the monthly average-per-square-foot for a
30-day month. For buildings that contain multiple accounts, we
subdivided the square footage evenly across all accounts.  For
example, Apartment Complex A, Building C, has an area of 4521 square
feet, and five gas accounts. Suppose the utility bill for one of these
five accounts from 2007-01-19 to 2007-02-20 (for 32 days) is 76
CCF.  %units of hundred cubic feet.
We computed its normalized consumption value for February of 2007 to
be: $30 \times \frac{76}{32 \times 4521/5} =
0.07880$. Figure~\ref{fig:datanorm}(a) shows the normalized data
measured in 30-day CCF per square foot.  This re-normalization
alleviates the issue of comparisons between months of different
lengths.

We aggregated the weather covariates across all days of the billing
period and then summarized within that period. Therefore, the weather
covariates may vary, even for observations with the same nominal
billing period, if the meters were observed across separate days.
Then, we additionally adjusted the observations by dividing the
normalized values by the sum of the HDD and CDD in the corresponding
month. We used these adjusted observations for our analysis.

\subsection{Missing Data Imputation}

Of the $238$ accounts, $39$ accounts have missing observations in
the time series interior. The number of missing values in any of the
$39$ time series was at most $7$, while $23$ series only had missing a
single value. We believe that it is reasonable to assume that these
values are missing at random, as they typically arise from clerical
oversight \citep{little2019statistical}.  We used the ``imputeTS''
package in R \citep{imputeTS} to impute the missing values in the
degree day-adjusted re-normalized observations.  Based on a visual
inspection of the output of the various interpolation, Kalman
smoothing, and moving average options in this package, we selected the
\textit{Kalman smoothing and structural time series model} option.  We
note that, if an imputed value was negative, we replaced the imputed
value with 0 (since consumption must be non-negative).  The
weather-adjusted re-normalized series with imputed values are shown in
Figure~\ref{fig:datanorm}(b).

\subsection{Grouping of Accounts}

Table~\ref{tab:acct-types} gives the number of accounts (each
corresponding to a monthly time series) for the seven utility
services.

\begin{table}[ht]
\centering
\begin{tabular}{llr}
  \toprule
  Utility Service Type & Service Group & \# of Accounts \\
  \midrule
  Non-Residential Small General Service & Small non-residential & 108 \\
  Non-Residential Medium General Service & Medium to large non-residential & 33  \\
  Non-Residential Large General Service & Medium to large non-residential & 1 \\
  Seasonal - Commercial & Medium to large non-residential & 1 \\
  Residential Heating & Residential & 74 \\
  Residential Multi-dwelling Large & Residential & 9 \\
  Residential Multi-dwelling Small & Residential & 1 \\
  \bottomrule
\end{tabular}
\caption{Frequency distribution of accounts by utility service
  types. By merging several service types, we formed three
  service groups.}
\label{tab:acct-types}
\end{table}

% Facilities Operations also provided classifications of the building's
% occupancy type, frequencies of which are given in in
% Table~\ref{tab:function}.

% \begin{table}[ht]
% \centering
% \begin{tabular}{lr}
%   \hline
%   Type & Count \\ 
%   \hline
%   Residential Facilities & 176 \\ 
%   Support Facilities &  14 \\ 
%   Office Facilities &  10 \\ 
%   Laboratory Facilities &   9 \\ 
%   Special Use Facilities &   8 \\ 
%   General Use Facilities &   7 \\ 
%   Academic/Admin Facilities &   6 \\ 
%   Classroom Facilities &   3 \\ 
%   Agriculture Barn &   2 \\ 
%   Academic/Support Facilities &   1 \\ 
%   Parking Garage &   1 \\ 
%    \hline
% \end{tabular}
% \caption{``Institutional Function'' building occupancy type count, by
%   account.}
% \label{tab:function}
% \end{table}

This classification reflects real-world groupings of the buildings and
accounts, including residential housing, academic and administrative
buildings, food service locations, and agricultural support
buildings. Different building use-types exhibit different consumption
profiles.  Broadly speaking, we expect accounts with shared physical
or functional characteristics to exhibit similar consumption based on
their use-types.  We use the service groups given in
Table~\ref{tab:acct-types} to form three homogeneous groups of
accounts: small non-residential; medium to large non-residential; and
residential.

\begin{figure}[H]
  \centering
  \includegraphics[width=\textwidth]{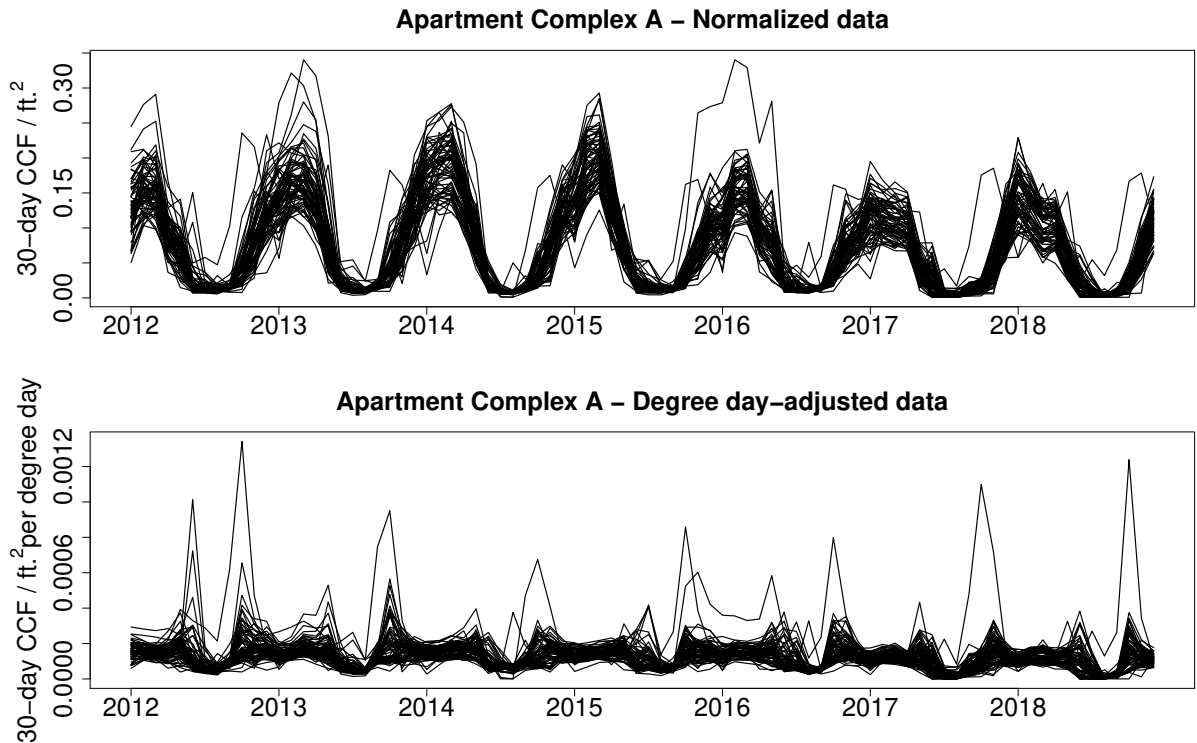}
  \caption{Historical data for $70$ gas accounts in Apartment
    Complex A. \emph{Top:} Normalized data, calculated as hundred
    cubic feet (CCF) per square foot, averaged over billing period
    duration and aggregated to a 30-day month. \emph{Bottom:}
    Degree day-adjusted and normalized data, calculated as per-degree
    day, where degree days are the sum of cooling and heating degree
    days.}
  \label{fig:mansfield}
\end{figure}

We illustrate this for undergraduate housing in Apartment Complex
A. Each unit occupies approximately 1000 square feet. Removing
accounts with missing values, we observe $N_1=70$ accounts. The
re-normalized data for these accounts are shown in the top of
Figure~\ref{fig:mansfield}.  The pattern of these time series reflects
the use of gas for heating in the winter. The accounts after degree
day adjustment are shown in the bottom panel of
Figure~\ref{fig:mansfield}.

We grouped the data into yearly cycles based on the university's
fiscal year, which begins on July 1 and continues until the following
June 30. This grouping coincides with both the university's accounting
schedule and the academic year. Student residents depart campus in
May, so logistically, summer is a good time for Facilities Operations
to address maintenance problems.

\section{Detecting Anomalies in Accounts from Known Homogeneous Groups} \label{sec:known-group}

\subsection{Model-Free Anomaly Detection on Adjusted Usage Data}\label{sec:model-free}

Anomaly detection could refer to identifying an account with behavior
that departs markedly from its own history, or from a normal pattern
that is expected based on known characteristics for that account.  For
example, the normal pattern for each month could be represented by a
median, or suitable quantiles of the expected usage of the accounts.

Based on observations from all $70$ accounts in Apartment Complex
A, we applied a moving window of two years and computed curves that
represent the 2.5\% and 97.5\% percentiles across both years. We then
considered using the curves as reference limits for normal behavior in
the following year. Our choice to use a two year window was motivated
by the high degree of variance we observed in the quantiles when
considering quantiles for only the single preceding year.

\begin{figure}[H]
  \centering
  \includegraphics[width=\textwidth]{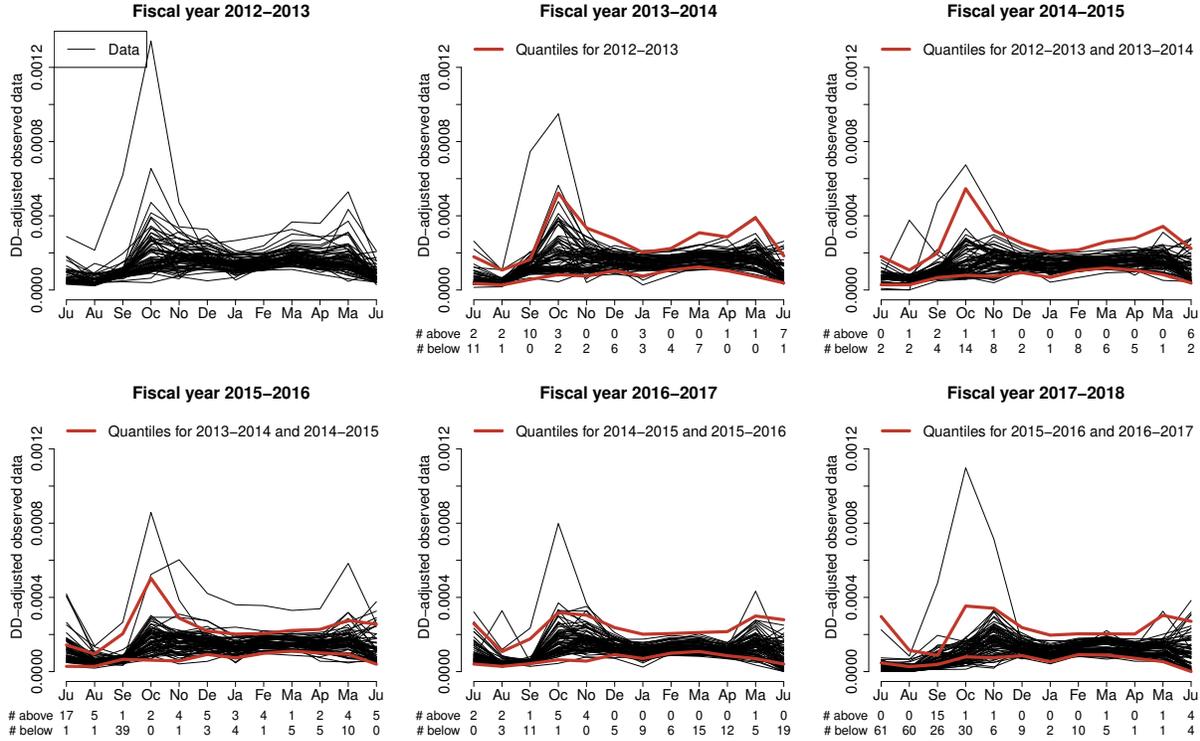}
  \caption{Fiscal year observations for Apartment Complex A. Red
    lines give previous two years' $2.5$\% and $97.5$\%
    quantiles. Below the chart is the number of accounts above the
    $97.5$\% quantile (line
    1) and below the $2.5$\% quantile (line 2).}
  \label{fig:years}
\end{figure}

We show six years (2012 to 2018) of weather-adjusted data in
Figure~\ref{fig:years}, and the quantiles for the previous two years
are shown in red.  Our intention was to use the quantiles to identify
accounts with an anomaly in some month, identified by a usage value
falling outside the reference quantiles. The two year window
accommodates systematic structural changes to the consumption pattern
over time. This tendency is apparent in the shift of peak consumption
from October in 2012--2013, to November in 2017--2018.

The reference quantiles derived from the previous two years represent
aggregates across multiple accounts.  For a specific building,
deviations from those reference quantiles will be due to idiosyncratic
factors of the new year: new occupants in a dormitory, new class
schedules in academic buildings. Therefore, they may be assumed to be
statistically independent of observations in the current year. We used
this to directly flag outlying values in. Below the horizontal axis of
each panel of Figure~\ref{fig:years}, we report the number of accounts
above and below the reference quantiles.

The reference quantiles captured the dominant central tendency of
recent observations. This permits identification of anomalous accounts
such as in 2015--2016, where a single account is consistently elevated
above other series. This demonstrates the importance of grouping
together buildings with similar consumption profiles: for a different
building use-type, that series might not be anomalous.  On the other
hand, we flag a large number of accounts as ``outliers'' from July to
October in 2017. This suggests that the accounts we flagged might not
provide a meaningful summary of deviation from a ``normal''
baseline. In particular, the past quantiles do not reflect behavior in the
current year.

Another important aspect of Figure~\ref{fig:years} is the recurrence
of level differences across accounts. This is apparent in the series
with repeatedly high usage in October. Because the magnitude of this
series dominates the other series, it is difficult to compare this
series with the others. This may reflect a systematic
difference in that account, but it also suggests that level shifts may
pose an obstacle to applying these quantiles to the data.

\subsection{Model-Free Anomaly Detection on Proportions}

We identified two problems with the quantile-based flagging, namely,
(1) inconsistent usage patterns from one year to the next, despite
within-group homogeneity in a single year; and (2) differences in
scale that may dominate usage patterns that are otherwise similar. To
remedy these problems, we calculated proportional usage for each month
within the fiscal year.  Denoting energy usage in a given year for
account $i$ in month $m$ by $x_{im}$, $i=1,\dots,N$, where $N=70$,
$m=1,\dots,12$, we computed the proportion $p_{im}$ of usage in month
$m$ relative to overall consumption:
\begin{gather}
p_{im} = x_{im} / \sum_{k=1}^{12} x_{ik}
\end{gather}
The resulting sequence is of length 12.

The proportion transformation permits us to compare accounts with
different magnitudes by removing level information about the series.
It also allows us to address the variability in usage patterns between
years, such as the shift of peak usage from October to November noted
above in Section~\ref{sec:model-free}. This is accomplished by summarizing the relative magnitude of
each month's observation, while also controlling for year-wide changes
in temperature allocation, occupancy of residences, and other factors
that may vary from year to year.

We considered applying Tukey's boxplot to the marginal monthly
proportions across all accounts. The boxplot defines a convenient
definition of an outlier, defined in terms of individual observations
from a homogeneous group. Denote the lower and upper quartiles of the
data (25\% and 75\% quantiles, respectively) by $Q_{1}$ and
$Q_{3}$. The interquartile range is defined as
$\text{IQR}=(Q_{3}-Q_{1})$, which is a robust estimate of the
population variance.

Tukey's thresholds for outliers are based on the lower and upper
fences, which are defined in Table~\ref{tab:bounds} for moderate and
severe outliers. The lower bound for non-outlier values is found as
the smallest observed value larger than the lower fence. Likewise, the
upper bound is the largest observed value smaller than the upper
fence.

\begin{table}[ht]
  \centering
  \begin{tabular}{lll}
    \toprule
    & Moderate outlier & Severe outlier \\
    \midrule
    Lower fence & $Q_{1} - 1.5\times\text{IQR}$ & $Q_{1} - 3\times\text{IQR}$ \\
    Upper fence & $Q_{3} + 1.5\times\text{IQR}$ & $Q_{3} + 3\times\text{IQR}$ \\
    \bottomrule
  \end{tabular}
  \caption{Lower and upper fences for outlier detection, based on
    Tukey's boxplot. The lower bound is the smallest value larger than
    the lower fence, and the upper bound is the largest value smaller
    than the upper fence. Observations below the lower bound or above
    the upper bound are flagged as outliers.}
  \label{tab:bounds}
\end{table}

\begin{figure}[H]
  \centering
  \includegraphics[width=.8\textwidth]{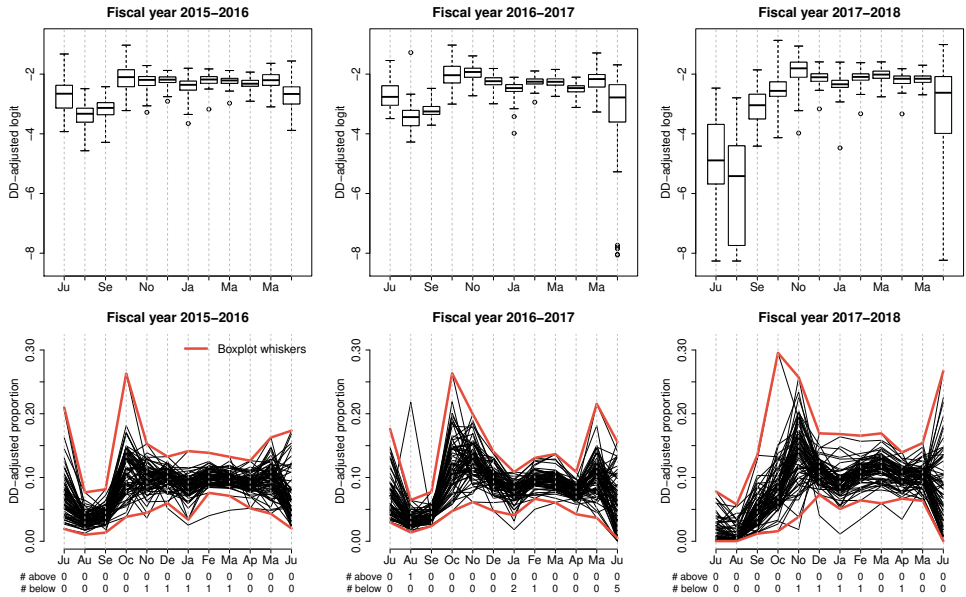}
  \caption{Boxplots on logit of monthly proportions of fiscal yearly
    consumption (top), and corresponding proportions (bottom).  Bounds
    corresponding to the boxplot whiskers are shown in red.}
  \label{fig:boxplot}
\end{figure}

These threshold bounds assume a normally-distributed sample, which is
clearly violated by proportion values. Therefore, we transformed each
proportion to the logit scale, defined by the transformation
\begin{gather}
  \label{eq:logit}
  \ell_{im}=\log\left(\frac{p_{im}}{1-p_{im}}\right)
\end{gather}
We then constructed boxplots within each month across all
accounts. These boxplots are shown for three fiscal years in the top
panel of Figure~\ref{fig:boxplot}.

\begin{figure}[H]
  \centering
  \includegraphics[width=.8\textwidth]{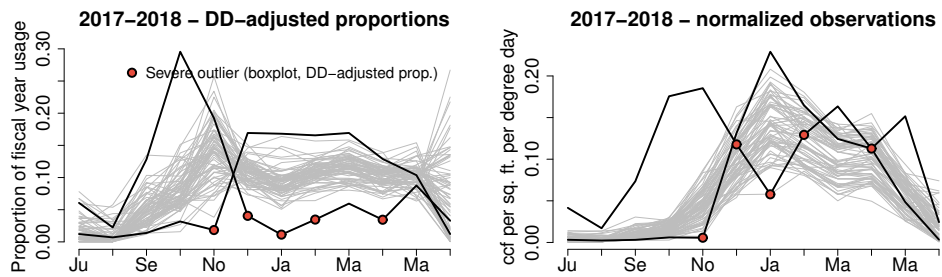}
  \caption{Outlier observations from 2017--2018. Proportion data is
    shown at left, degree-day-adjusted data on the right. Series with
    outliers are highlighted, and outlier observations are marked.}
  \label{fig:outliers}
\end{figure}

The number of outliers is marked below the bottom panel of
Figure~\ref{fig:boxplot}. The boxplots identify two of the residential
accounts as outliers in fiscal year 2017--2018.  The proportions for
these series are highlighted in Figure~\ref{fig:outliers}.  One of the
accounts exhibits elevated usage in October relative to other accounts
in the group, and relatively low usage from December to April. The
decrease exhibits the outliers, although the specific decreases in the
proportions may be due to a linear response to the exceptionally large
usage in October.  The other series exhibits low usage in November,
and relatively high usage from December to March. The outlier data
point itself is the low usage in November, and the elevated usage in
the winter does not register as statisticall significant.  The
proportions clealry demonstrate lower
consumption in some months, although these same points appear far less
abnormal in the non-proportion data.

\begin{figure}[H]
  \centering
  \includegraphics[width=.8\textwidth]{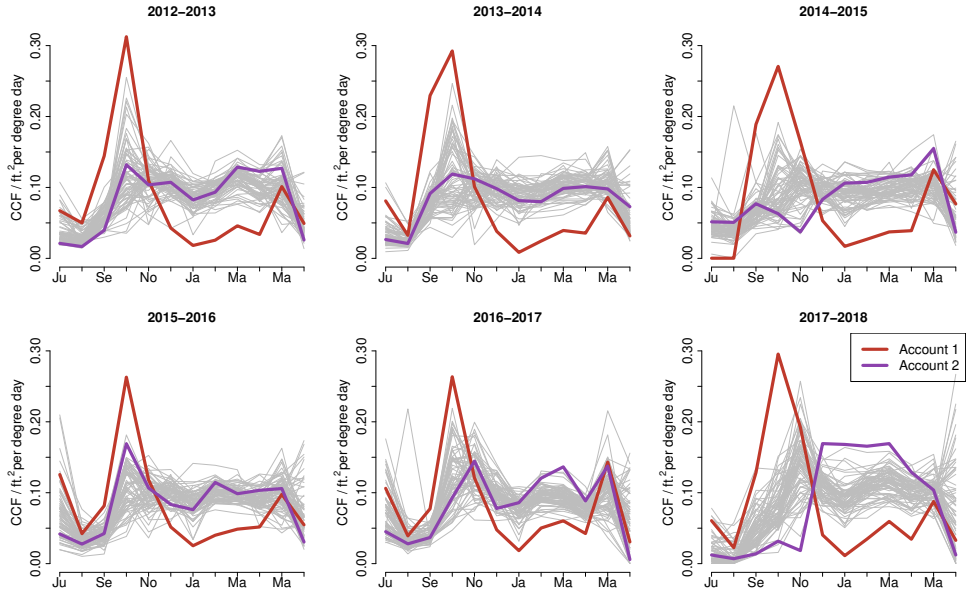}
  \caption{Outlier time series from fiscal year 2017--2018.}
  \label{fig:1718}
\end{figure}

We follow in detail the historical paths of these two accounts in
Figure~\ref{fig:1718}.  In previous years, we observe that the
behavior of Account 2 is comparable to the other accounts in Apartment
Complex A, with a marked change in 2017--2018. In contrast, the large
value in Account 1 recurs with the same pattern across years, and is
consistently out-of-sync with the other buildings.  Further
investigation revealed that Account 1 is the only account in Apartment
Complex A that is classified by the utility as ``non-residential small
general service'', instead of ``residential heating.'' Moreover, a
Facilities Operations annotation for this account indicates this account is used
for the laundry dryer in Building C.

Therefore, we understand Account 1 to represent behavior that is truly
outlying, relative to the other accounts, because we may attribute
differences in consumption to known external factors. We thus
conjecture that there may be a real anomaly in Account 2, as well,
because of incompatibility with its own past behavior, as well as the
extreme low consumption in November.  Further investigation by
Facilities Operations will clarify the status of this account, and
assess the reality of any anomaly there.

\subsection{Model-Based Anomaly Detection on Proportions} \label{sec:model}

In light of the lack of publications that consider anomaly detection
on proporitions, we propose a linear model that offers an approach to
monitor each new observation as it arrives, in an online fashion.  We
adapt the general monitoring framework of \citet{fu2014spc}, which
implements hypothesis tests to monitor for mean changes of a given
magnitude.  These authors developed an integrated likelihood ratio
(ILR) test to implement a procedure similar to Bartlett's sequential
probability ratio test (SPRT).  Since a``Bartlett-type likelihood
ratio'' (BLR) test would require multivariate integration across many
dimensions as well as a maximization over nuisance parameters, they
instead used historical in-control data to estimate the nuisance
parameters, leading to an approximate Bartlett-type likelihood ratio
(ABLR).

As above, we consider observations of monthly energy consumption data
for N utility accounts. We further consider $K$ historical years of
data, and modify our notation accordingly. Denote the consumption in
month $m$ of year $k$ for the $i$th account by $x_{ikm}$,
$i=1,\dots,N$, $k=1,\dots,K$, and $m=1,\dots,12$. The dataset contains
$12KN$ observations.

Each year of data is a 12-month cycle for each account. For each month
$m$ in year $k$ for the $i$th account, we calculate the proportion of
overall yearly consumption by
\begin{gather}
  p_{ikm} = \frac{x_{ikm}}{\sum_{\ell=1}^{12}x_{ik\ell}}.
\end{gather}

Finally, we apply the logistic transformation to $p_{ikm}$,
$m=1,\dots,12$, yielding data
\begin{gather}
  \label{eq:logity}
  y_{ikm} = \log \left( \frac{p_{ikm}}{1-p_{ikm}} \right).
\end{gather}

Let $\y_{ik}=(y_{ik1},\dots,y_{ik12})'$ denote the 12-dimensional vector
of logit-transformed, yearly proportional consumptions.  We assume
\begin{gather}
  y_{ik} = \bbeta_i + \bepsilon_{ik},  \\
  \bepsilon_{ik}=(\epsilon_{ik1},\dots,\epsilon_{ik,12})' \sim N(\zero, \Sigma),
\end{gather}
where $\bbeta_i=(\beta_{i1},\dots, \beta_{i,12})'$ is the mean of the logit transformed yearly proportional consumptions for the $i$th account for $i=1,\dots,N$.

Let $D_K=\{\y_{ik}, i=1,\dots,N, k=1,\dots,K\}$ denote the ``historical'' data for these $N$ accounts.
Write $\bbeta=(\bbeta_1',\dots,\bbeta_N')'$.
The likelihood function based on $D_K$ is
\begin{equation}
  {\cal L}(\bbeta, \Sigma|D_K)
  \propto |\Sigma|^{-KN/2}\exp\left\{ - \frac{1}{2} \sum^K_{k=1}
  \sum_{i=1}^{N} (\y_{ik}-\bbeta_i)'\Sigma^{-1}(\y_{ik}-\bbeta_i)  \right\}.
    \label{histlike}
\end{equation}
Then, the maximum likelihood estimates of $\bbeta_1, \dots,\bbeta_N$, and $ \Sigma$ are
\begin{gather}
  \hat{\bbeta}_{i} = \frac{1}{K}\sum_{k=1}^{K}\y_{ik}, \;\; i=1,\dots, N \\
  \widehat{\Sigma} = \frac{1}{NK}\sum_{i=1}^{N}\sum_{k=1}^{K} (\y_{ik}-\hat{\bbeta}_{i})(\y_{ik}-\hat{\bbeta}_{i})'.
\end{gather}

Let $\y_{K+1}=(\y_{1,K+1}',\dots,\y_{N,K+1}')'$ denote the logit
transformed yearly proportional consumptions of $N$ accounts at year
$K+1$.  Let $C=\mbox{diag}(c_1,c_2,\dots,c_{12})$ be a fixed
$12 \times 12$ diagonal matrix such that $C \ne I_{12}$, where
$I_{12}$ is the $12 \times 12$ identity matrix. Here the values of the
diagonal elements of $C$ specify the magnitudes of change in
$\bbeta_i$ that indicate the underlying consumption is no longer
in-control for the $i$th account. In other words, we test the
hypothesis
\begin{gather}
  H_{0}: \bbeta_i=\bbeta^{(0)}_i\quad\text{vs.}\quad H_{1}: \bbeta_i=C\bbeta^{(0)}_i
\end{gather}
as we observe the new value of $\y_{i,K+1}$ for $i=1,\dots,N$.

The test statistic is
\begin{equation*}
  T_{i}^{ABLR}=\frac{ \exp\Big\{- \frac{1}{2} (\y_{i,K+1} - C\bbeta^{(0)}_i)'\Sigma^{-1} (\y_{i,K+1} - C\bbeta^{(0)}_i)\Big\} }
   {\exp\Big\{- \frac{1}{2} (\y_{i,K+1} - \bbeta^{(0)}_i)'\Sigma^{-1} (\y_{i,K+1} - \bbeta^{(0)}_i)\Big\}   }
\end{equation*}
and
\begin{align*}
  \log \Big(T_{i}^{ABLR} \Big)= &  ( \bbeta^{(0)}_i)'(C-I_{12}) \Sigma^{-1}\y_{i,K+1}
   - \frac{1}{2}( \bbeta^{(0)}_i)C \Sigma^{-1}C \bbeta^{(0)}_i
    + \frac{1}{2}( \bbeta^{(0)}_i) \Sigma^{-1} \bbeta^{(0)}_i.
\end{align*}
Under $H_0$,  we have
\begin{align}
   & \log \Big(T_{i}^{ABLR} \Big)  \nonumber \\
 \sim &
  N\Big( (\bbeta^{(0)}_i)'(C-I_{12}) \Sigma^{-1}\bbeta^{(0)}_i  - \frac{1}{2}( \bbeta^{(0)}_i)C \Sigma^{-1}C \bbeta^{(0)}_i
    + \frac{1}{2}( \bbeta^{(0)}_i) \Sigma^{-1} \bbeta^{(0)}_i ,  \nonumber  \\
  & \;\;\;\;    (\bbeta^{(0)}_i)'(C-I_{12}) \Sigma^{-1} (C-I_{12}) \bbeta^{(0)}_i
  \Big).
\end{align}
Write
\begin{equation}
Z_i(\y_{i,K+1}, \bbeta^{(0)}_i, \Sigma) = \frac{ \Big| ( \bbeta^{(0)}_i)'(C-I_{12}) \Sigma^{-1}\y_{i,K+1}
  - (\bbeta^{(0)}_i)'(C-I_{12}) \Sigma^{-1}\bbeta^{(0)}_i \Big| }
{ \Big\{ (\bbeta^{(0)}_i)'(C-I_{12}) \Sigma^{-1} (C-I_{12}) \bbeta^{(0)}_i \Big\}^{1/2} }.
\end{equation}
Let
\begin{equation}
Z_{\max}(\y_{K+1}, \bbeta^{(0)}, \Sigma)= \max \{ Z_1(\y_{1,K+1}, \bbeta^{(0)}_1, \Sigma),\dots,Z_N(\y_{N,K+1}, \bbeta^{(0)}_N, \Sigma) \},
\end{equation}
where $ \bbeta^{(0)}=( (\bbeta^{(0)}_1)', \dots, (\bbeta^{(0)}_N)')'$.
Note that under $H_0$, we have
$$
\frac{ ( \bbeta^{(0)}_i)'(C-I_{12}) \Sigma^{-1}\y_{i,K+1}
  - (\bbeta^{(0)}_i)'(C-I_{12}) \Sigma^{-1}\bbeta^{(0)}_i  }
{ \Big\{ (\bbeta^{(0)}_i)'(C-I_{12}) \Sigma^{-1} (C-I_{12}) \bbeta^{(0)}_i \Big\}^{1/2} }
\sim  N(0,1)
$$
 for $i=1,\dots,N$. Thus, under $H_0$ and for $z>0$, we have
$$
P(Z_{\max}(\y_{K+1}, \bbeta^{(0)}, \Sigma)  < z ) = [2 \Phi(z)-1]^N,
$$
where $\Phi(\cdot)$ is the standard normal cumulative distribution function.
For a given significance level $\alpha$, the rejection region is given by
\begin{equation}
\{  Z_{\max}(\y_{K+1}, \bbeta_0, \Sigma)  \geq  z_{ (1+\sqrt[N]{1-\alpha})/2} \},
\end{equation}
where  $z_{(1+ \sqrt[N]{1-\alpha})/2 }$ is the $((1+\sqrt[N]{1-\alpha})/2)$th quantile of $N(0,1)$, i.e.,
$ \Phi(  z_{ (1+\sqrt[N]{1-\alpha})/2} ) = (1+\sqrt[N]{1-\alpha})/2 $.

Because the number of years $K$ is generally small, we also seek to
account for the variability in the estimates of $\bbeta_{i}$ and
$\Sigma$. Therefore, we adopt a noninformative Jeffrey's-type prior
distribution, namely,
\begin{gather}
  \pi(\bbeta, \Sigma) \propto |\Sigma|^{-1/2}
\end{gather}
In combination with the likelihood, this gives the posterior distribution
\begin{gather}
  \pi(\bbeta, \Sigma|D_{K})\propto \Sigma^{-(KN+1)/2}\exp
  \left\{-\frac{1}{2}
    \sum_{k=1}^{K}\sum_{i=1}^{N}(\y_{ik}-\bbeta)'\Sigma^{-1}(\y_{ik}-\bbeta_{i})
  \right\}
\end{gather}
We sample from this sequentially, with
\begin{gather}
  \bbeta_{i}|\Sigma \sim N(\hat\bbeta_{i}, \Sigma/K), \enspace
  i=1,\dots,N \\
  \Sigma^{-1} \sim {\cal W}(\Psi^{-1}, NK-12) \\
  \Psi = \sum_{k=1}^{K}\sum_{i=1}^{N}(\y_{ik}-\bbeta_i)(\y_{ik}-\bbeta_i)’
\end{gather}
where ${\cal W}(\Psi, \nu)$ is a Wishart distribution with scale
matrix $\Psi$ and $\nu$ degrees of freedom.

This yields the following algorithm for online monitoring of the
proportional usage data.
\noindent {\bf Yearly Monitoring Algorithm}
\begin{description}
\item[Step 0] Set significance level $\alpha$
\item[Step 1] Compute credible level $\gamma_{\alpha}$
  \begin{itemize}
  \item For each of $M$ replicates, perform the following simulations:
    \begin{itemize}
    \item Simulate new data $\y_{i,K+1}^{*(m)} \sim N(\hat{\bbeta}_i,
      \widehat{\Sigma})$ independently for $i=1,\dots,N$
    \item Perform $B_1$ replicates of the following simulation:

    (i) Generate $\bbeta^{(0)}_{mb}$ and $\Sigma_{mb}$ from the ``posterior'' distribution
      \begin{equation}
      \pi(\bbeta^{(0)},\Sigma|D_{K})
  \propto \Sigma^{-(KN+1)/2}\exp\left\{ - \frac{1}{2} \sum^K_{k=1}
  \sum_{i=1}^{N} (\y_{ik}-\bbeta^{(0)}_i)'\Sigma^{-1}(\y_{ik}-\bbeta^{(0)}_i)  \right\}.
    \end{equation}
   (ii) Calculate the indicator function
        \begin{gather}
          \delta_{b}(\y^{*(m)}_{K+1},\bbeta^{(0)}_{mb}, \Sigma_{mb}) =
          \mathbbm{1} \left\{  Z_{\max}(\y^{*(m)}_{K+1}, \bbeta^{(0)}_{mb}, \Sigma_{mb})
            \geq z_{ (1+\sqrt[N]{1-\alpha})/2}  \right\},
        \end{gather}
        where $ \y^{*(m)}=( (\y_{1,K+1}^{*(m)})',\dots, \y_{N,K+1}^{*(m)})'$
        for $b=1,\dots,B_1$.

    \item Calculate $\hat q_{m} =\frac{1}{B_1} \sum_{b=1}^{B_1}\delta_{b}(\y^{*(m)}_{K+1},\bbeta^{(0)}_{mb}, \Sigma_{mb} )$.
    \end{itemize}
    for $m=1,\dots, M$.
  \item Compute $\gamma_{\alpha}$ as the $\alpha$th percentile of the
    empirical distribution of the $M$ values of $\hat q_{m}$'s.
  \end{itemize}

\item[Step 2] Monitor
  \begin{itemize}
  \item Using the observation $\y_{K+1}$ in the $K+1$ year, perform $B_2$
    replicates of the following:

     (i) Generate $\bbeta^{(0)}_{b}$ and $\Sigma_{b}$ from the ``posterior'' distribution
      \begin{equation}
      \pi(\bbeta^{(0)},\Sigma|D_{K})
  \propto \Sigma^{-(KN+1)/2}\exp\left\{ - \frac{1}{2} \sum^K_{k=1}
  \sum_{i=1}^{N} (\y_{ik}-\bbeta^{(0)}_i)'\Sigma^{-1}(\y_{ik}-\bbeta^{(0)}_i)  \right\}.
    \end{equation}
   (ii) Calculate the indicator function
        \begin{gather}
          \delta_{b}(\y_{K+1},\bbeta^{(0)}_{b}, \Sigma_{b}) =
          \mathbbm{1} \left\{  Z_{\max}(\y_{K+1}, \bbeta^{(0)}_{b}, \Sigma_{b})
            \geq z_{ (1+\sqrt[N]{1-\alpha})/2}  \right\}
        \end{gather}
        for $b=1,\dots,B_2$.

    \item Calculate $\hat q =\frac{1}{B_2} \sum_{b=1}^{B_2}\delta_{b}(\y_{K+1},\bbeta^{(0)}_{b}, \Sigma_{b})$.

    \item If $\hat q \geq \gamma_{\alpha}$, do not reject $H_{0}$, i.e., no abnormal consumptions in the $K+1$ year.

    \item If $\hat q < \gamma_{\alpha}$, reject $H_{0}$, and report that there are abnormal logit transformed yearly proportional consumptions
    for at least one account, which corresponds to account $i^\ast$ such that
     $$
     i^\ast = \mbox{argmax}_{1 \leq i \leq N} \{ Z_i(\y_{i,K+1}, \widehat{\bbeta}_i, \widehat{\Sigma}) \}.
     $$

  \end{itemize}

\end{description}

\subsection{Comparison of Model-Free and Model-Based Monitoring
  Procedures}

A feature of the transformation in Equation~\ref{eq:logity} is that it
applies separately to each of the 12 months of the year. This
contrasts with the more traditional multiple logistic transformation
that would express each of 11 months relative to a 12th, benchmark
month, which omits one month in order to enforce the constraint that
the 12 proportions sum to unity.  As we observe in the boxplot-based
method, a substantial shift in the consumption of a single month
affects the proportion values for the other 11 months, as well, which
introduces complicated dynamics in the relative behavior of the
overall proportion vector. Expressed in the diagonal matrix $C$, this
chaotic behavior would require search over a large space of possible
values of $C$ to elaborate the wide range of possible distortions
across the 12 monthly proportions.

Our method is far simpler, and far more interpretable: for a given
month $m$ in a given account $i$, the alternative hypothesis specifies
a multiplicative increase in the log odds of the proportion $p_{ikm}$.
A disadvantage of this approach is that it is not the case that the
inverse-logit transformed proportions sum to 1, but the benefit is
interpretibility of the change in a given month's
proportion. Additionally, this approach also sidesteps the possibility
for chaotic behavior across the full proportion vector $\y_{ik}$, in
favor of a marginal effect within each month that does not directly
impact the mean value in other months.

To demonstrate the relative performance of the two monitoring
approaches, we proceed with a constructive data analysis. We start
with the accounts 1 and 2, identified by the model-free procedure as
outliers. We also estimate the full statistical model across $K=5$
years.

Suppose we know \emph{a priori} that a given series $i$ is an outlier in the
sense of being out-of-control in a given year. Within the model, this
means, the mean value of the new year $(K+1)$ is $C\beta_i$.
Under $H_{1}$, we can calculate the effective nominal increase in
magnitude for a given month $m$ as $\hat{C}_{im}=y_{i(K+1)m} /
\beta_{im}$. The the accounts with
large values along the diagonal of $\hat{C}$ are of intuitive
interest, so we consider two more accounts. First, we calculate the
sum of the squared diagonal for the full matrix $\hat{C}_{i}$, namely,
$\sum_{m=1}^{12} \hat{C}_{im}^2$, and consider the account with the
large such value. For the second account, we apply the same procedure
to the subset of indices $m$ that correspond to the months
October--April.

\begin{figure}[H]
  \centering
  \includegraphics[width=\textwidth]{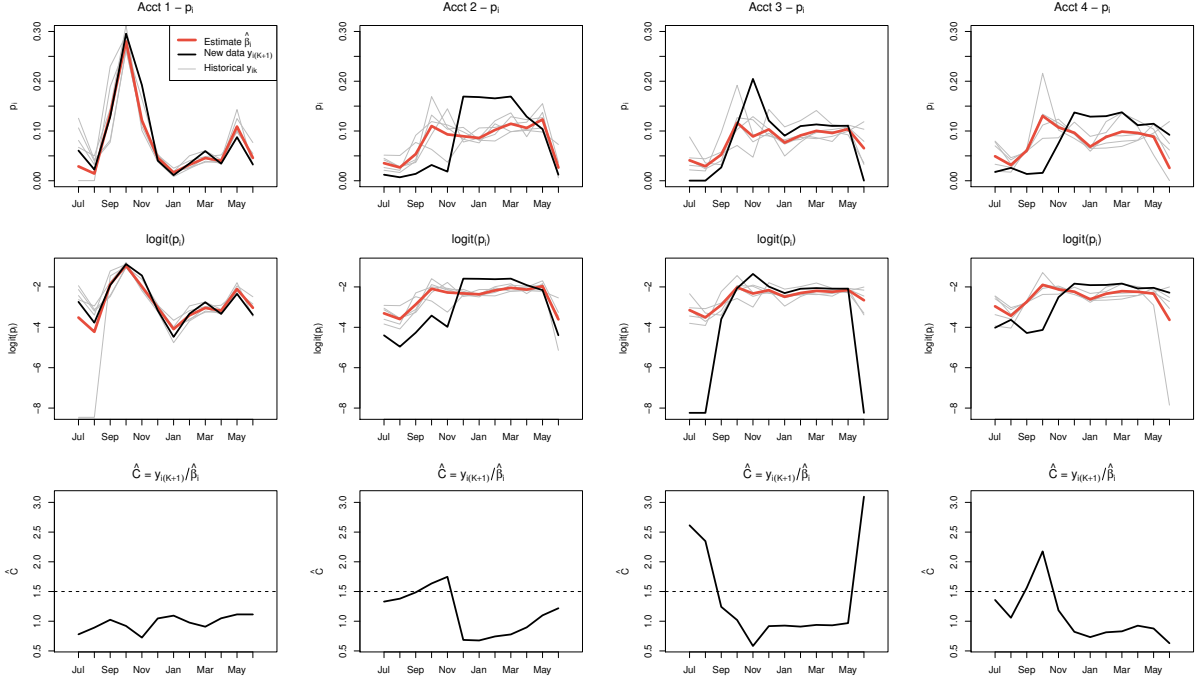}
  \caption{\emph{Top:} Monthly proportional energy consumption $p_{i}$ for
    four accounts, accounts 1 and 2 identified using the model-less
    approach, and accounts 3 and 4 identified as exhibiting the
    largest ratio
    $\hat{C}_{i}=\sum_{m=1}^{12}y_{i(K+1)m}/\hat{\beta}_{im}$ across
    all the data (account 3) and across only the months October--April
    (account 4). Shown are in-control
    historical proportions and logit-transformed mean
    value. Hypothesis testing is performed for the single new
    observation in period $(K+1)$, or 2017-2018.
  \emph{Middle:} Monthly logit-scale proportional usage values, showing
  estimated mean value and corresponding transformed data $y_{i}$ for
  each account. Estimation and hypothesis tests are performed on these
  transformed data points.
  \emph{Bottom:} Empirical increase in monthly consupmtion in year
  $(K+1)$ as a percent of the mean value. Noted is the threshold
  $C=1.5$, which reliably identifies large deviations from the mean value.}
  \label{fig:logits}
\end{figure}

These four accounts are shown in Figure~\ref{fig:logits}. In the top
panel, we plot the fitted mean value based on $\hat\bbeta_{i}$ on the
proportional scale, as well as the new data observation $\y_{i}$.  For
reference, the historical data is shown in gray.  The second row shows
the same data, transformed to the logit scale.  Finally, the bottom
row shows the ratio of $\y_{i(K+1)}$ to $\hat\bbeta_{i}$, which can be
thought of as the nominal proportional increase $\hat {\mathbf{C}}_{i}$
of the new observation to its mean, the most direct estimate of the
true change from the in-control value.

As noted, the first two accounts in Figure~\ref{fig:logits} were
identified using the model-free procedure. Two scenarios are present:
the new observation in period $(K+1)$ is consistent with past
behavior, as in the first account, and the model-free method likely
identified the outlier based on deviations from the group-wide
trends. In the second account, the observation in the new period
exhibits substantial differences from the historical
trend. Accordingly, the first example is not likely to be well-suited
to identification with the model-based approach.

The figure also shows the transformed values of the proportions on the
logit scale, as well as the empirical estimates of $\hat{C}$ for both
accounts. For the first account, we see a stable ratio of observed,
new values to the estimates based on historical data, and none of the
ratios is much larger than 1. On the other hand, the second account
exhibits a sharp increase in the mean parameter associated with
November. The magnitude is large at over 1.5, and this is an
intuitively appealing value. Moreover, the second is of the type of
outlier the model is well-suited to identify: the new observation is
not consistent with past behavior during the cold months, from
November to March. The decrease in November may reflect the increase
in December, a nonlinear relationship that is difficult to specify in
terms of the matrix $\mathbf{C}$. Nonetheless, the empirical estimates
suggest a clear discontinuity between the current observation and
estimates based on its past behavior.

We do note that the first outlier account exhibits a one historical
series with low numerical stability: in one year, July and August both
exhibited very low consumption in that account, and this is
exaggerated by the logit transformation into a very large, negative
value. The estimation procedure itself is robust to this large
observation, and the estimate for that month is only somewhat
influenced by these outliers.

In the first of the remaining two accounts, we observed large values
of $\hat{\mathbf{C}}_{i}$, but they occur in the edge months of July, August, and
June. Although the magnitude is substantial, the real proportion of
usage is quite small in these months. Although the differences may be
statistically large, these deviations are not practically
significant—they do not appear to reflect major behavioral changes in
the series' yearly proportional pattern, so much as artifacts of the
logit transformation. This suggests that, in order for the model-based
approach to work, we should monitor for changes of magnitude around
$c=1.5$ in magnitude,
only in the winter months.

Therefore, we can draw two conclusions on the basis of the visual
representation. First, the model-based procedure can capture
significant, out-of-control behavior, in terms of the estimated ratio
$\hat{C}$. However, the types of out-of-control behavior come with
caveats: we require an explicit magnitude $C$ of change in the mean
parameters $\beta_i$, we should only consider specific months, and the
we must calibrate the values of all 12 diagonal elements $C_m$ to
accommodate a change of specific magnitude in only one month. Similar
exercises are necessary for changes in each of the other months that
may be of interest.

Second, we observe that the model-based approach has low power to
identify patterns in the proportions wherein a single account exhibits
behavior out-of-sync with other, similar accounts. The grouping
structure is only employed in the model-based approach to account for
low sample sizes, and pool estimation of the variance. The model
itself does not capture deviations in single accounts from their
group-wise trends.

Despite its comparative simplicity, the model-free approach requires
far fewer assumptions, and identifies a wider range of out-of-control
behaviors in the data.

Because of small sample sizes, the model-based approach requires more
detailed specification of the full scenarios, and the grouping of
similar accounts is a necessary component of the model to permit
pooling of information.  But, this will only increase precision on
estimates of individual accounts' mean profiles, which may miss
accounts that deviate from group-wise patterns not captured by an
account-level model.

\section{Statistically Clustering to Identify Homogeneous Groups
  Groups} \label{sec:cluster}

For accounts such as those in Apartment Complex A, the homogeneous
behavior we observe in the data reinforces known structural
similarities between the accounts. But, in cases such as the single
account of service type ``seasonal - commercial'' in
Table~\ref{tab:acct-types}, such a grouping is unavailable. To this
end, we also sought a statistical method to group accounts with
similar behavior. This would be followed by flagging anomalous
accounts and / or times using methods similar to
Section~\ref{sec:known-group}.  We describe hierarchical
clustering based on the intra-year proportions introduced in Section
\ref{sec:known-group}.

We constructed fiscal year proportions for $233$ accounts,
removing one account with a negative observation; three accounts with
missing observations; and one account in Apartment Complex B that is
primarily 0. We considered two years of data, fiscal years
2016--2018.  We computed the Euclidean distances between accounts
based on 24 data points, under the assumption they are
independent. For two vectors of 24 proportions (two years of
proportions), $\mathbf{p}_{i}$ and $\mathbf{p}_{j}$, $i\neq j$, we
calculated the distances
$d(\mathbf{p}_{i},\mathbf{p}_{j})=\sqrt{\sum_{k=1}^{24}(p_{ik}-p_{jk})^{2}}$

We then applied hierarchical clustering \citep{johnson2002applied}. We
used Ward's method to cluster the accounts
\citep{ward1963hierarchical}, which minimizes the error sum-of-squares
when agglomerating clusters.  To select the number of clusters, we
used the silhouette criterion. This offers a
conventional procedure for identifying the number of clusters in a set
of data by using the tightness of groups of observations to determine the optimal number of
groups \citep{rousseeuw1987silhouettes}.

\begin{figure}[H]
  \centering
  \includegraphics[width=.8\textwidth]{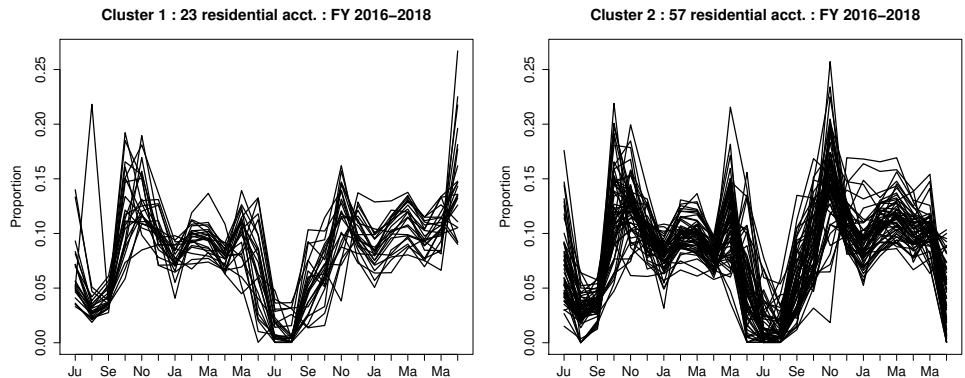}
  \caption{Fiscal-yearly proportion vectors for clustered residential
    accounts (Apartment Complexes A and B).}
  \label{fig:res-clusters}
\end{figure}

We first applied the clustering to the residential utility accounts,
as defined in Section~\ref{sec:data-desc} but with some series
removed, as previously discussed in this section. We considered 69
accounts in Apartment Complex A and 11 accounts in Apartment Complex
B. The clusters are shown in Figure~\ref{fig:res-clusters}. In
particular, cluster 1 consists of 23 accounts from Apartment Complex
A, while cluster 2 contains all of Apartment Complex B accounts and
the remaining 46 accounts in Complex A. The clusters are distinguished
by comparatively high usage in June in cluster 1 and a smoother
overall pattern during the academic year, and high November usage in
cluster 2 with a greater decrease in January.

To cluster the non-residential accounts, we used the service groups
defined in Table~\ref{tab:acct-types}, namely, small and
medium-to-large accounts. Within each of these groups, we performed
hierarchical clustering. We combined the partitions of all
non-residential accounts, defined in terms of the account size and the
two clusterings. Finally, we combined all degenerate singleton
clusters.

The medium-to-large accounts were found to have seven clusters, two of
which were degenerate, and the small accounts contained three
clusters, one of which was degenerate. This yielded the final eight
clusters shown in Figure~\ref{fig:res-clusters}. In addition to the
visual homogeneity of these groups, we report in
Table~\ref{tab:nonres-clusters} the cluster assignment counts for all
non-residential accounts for buildings containing more than one
account. We note the grouping of accounts according to the known building
labels, in general, such as Apartment Complex F and Apartment Complex
E, Building 4.

\begin{figure}[H]
  \centering
  \includegraphics[width=\textwidth]{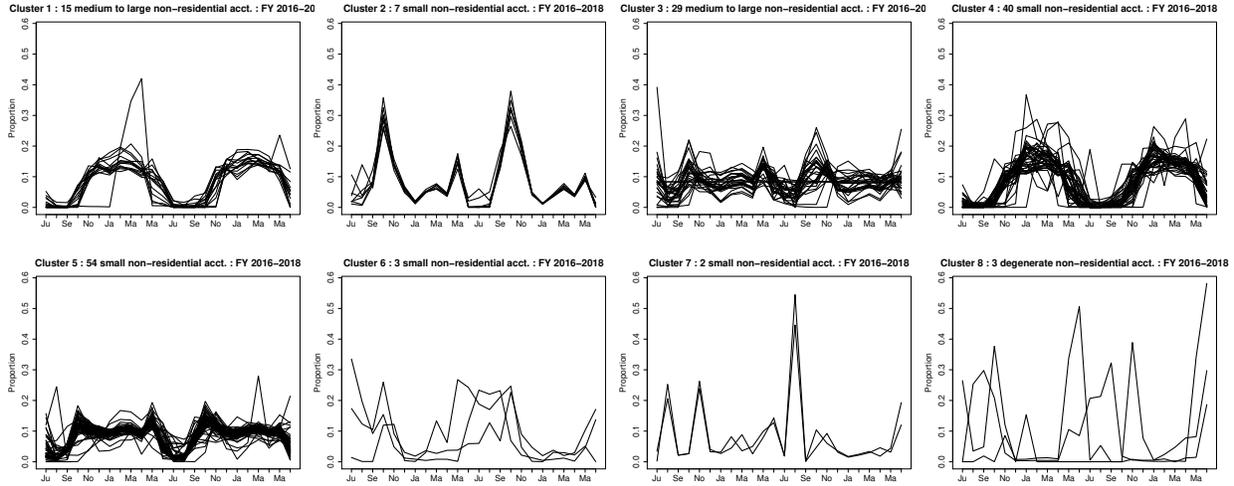}
  \caption{Cluster partition of utility-designated ``non-residential'' accounts.}
  \label{fig:nonres-clusters}
\end{figure}

%% Beef = Ag 1
%% Horse Barn = Ag 2
%% Sprague Hall = Dormitory 1
%% Ellsworth Hall = Dormitory 2
%% Foster Hall = Dorm 3
%% Hilltop Apt = Apartment Complex C
%% Husky Village = APt Complex D
%% Hough Hall = Apt Complex E, Bldg 1
%% Hubbard Hall = Apt Complex E, Bldg 2
%% Thompson Hall = Apartment Complex E, Bldg 3
%% Hoisington Hall = Apartment Complex E, Bldg 4
%% Busby Suites = Apartment Complex F

\begin{table}[ht]
\centering
\begin{tabular}{rrrrrrrrr}
  \toprule
  & 1 & 2 & 3 & 4 & 5 & 6 & 7 & 8 \\
  \midrule
  Ag 1 &   - &   - &   - &   1 &   1 &   - &   - &   - \\
  Ag 2 &   1 &   - &   - &   1 &   - &   - &   - &   - \\
  Dormitory 1 &   - &   - &   - &   - &   - &   1 &   1 &   - \\
  Student Union &   - &   - &   2 &   - &   - &   - &   - &   - \\
  Dormitory 2 &   - &   - &   2 &   - &   1 &   - &   - &   - \\
  Dormitory 3 &   - &   - &   - &   - &   7 &   - &   - &   - \\
  Apartment Complex C &   - &   - &   7 &   - &   - &   - &   - &   - \\
  Apartment Complex D &   - &   - &   4 &   - &   3 &   - &   - &   - \\
  Apartment Complex E, Bldg. 1 &   - &   - &   - &   - &   8 &   - &   - &   - \\
  Apartment Complex E, Bldg. 2 &   - &   - &   - &   1 &   7 &   - &   - &   - \\
  Apartment Complex E, Bldg. 3 &   - &   - &   - &   - &   8 &   - &   - &   - \\
  Apartment Complex E, Bldg. 4 &   - &   - &   - &   - &  12 &   - &   - &   - \\
  Apartment Complex F &   - &   - &   1 &  20 &   - &   - &   - &   - \\
  \bottomrule
\end{tabular}
\caption{Cluster assignment for utility-designated ``non-residential''
  accounts.}
\label{tab:nonres-clusters}
\end{table}

Within each of the homogeneous clusters, we then used the approach of
Section~\ref{sec:known-group}. Under this procedure, we identify
possible anomalies across all accounts, even those for which known
homogeneous structure is unavailable.

\section{Implementation for Management} \label{sec:implementation}

Having specified a method for anomaly detection among fiscal yearly
proportions, the final steps are implementation, integration into the
Facilities Operations energy management workflow, and a mechanism to incorporate
feedback from engineers into the statistical model.
We pursued seamless interaction between the Department of Statistics
and Facilities Operations through an intense collaboration on a
weekly, or even more frequent, basis. These meetings occurred
face-to-face as well as electronically, and provided opportunities for
dynamic exchange of ideas.

Implementation requires, on the one hand, aggregation, storage, and
analysis of the data; and on the other hand, an interface with
which Facilities Operations may review the results of the anomaly
detection as well as manage the information provided by the analysis.

\begin{figure}[H]
  \centering
  \includegraphics[width=\textwidth]{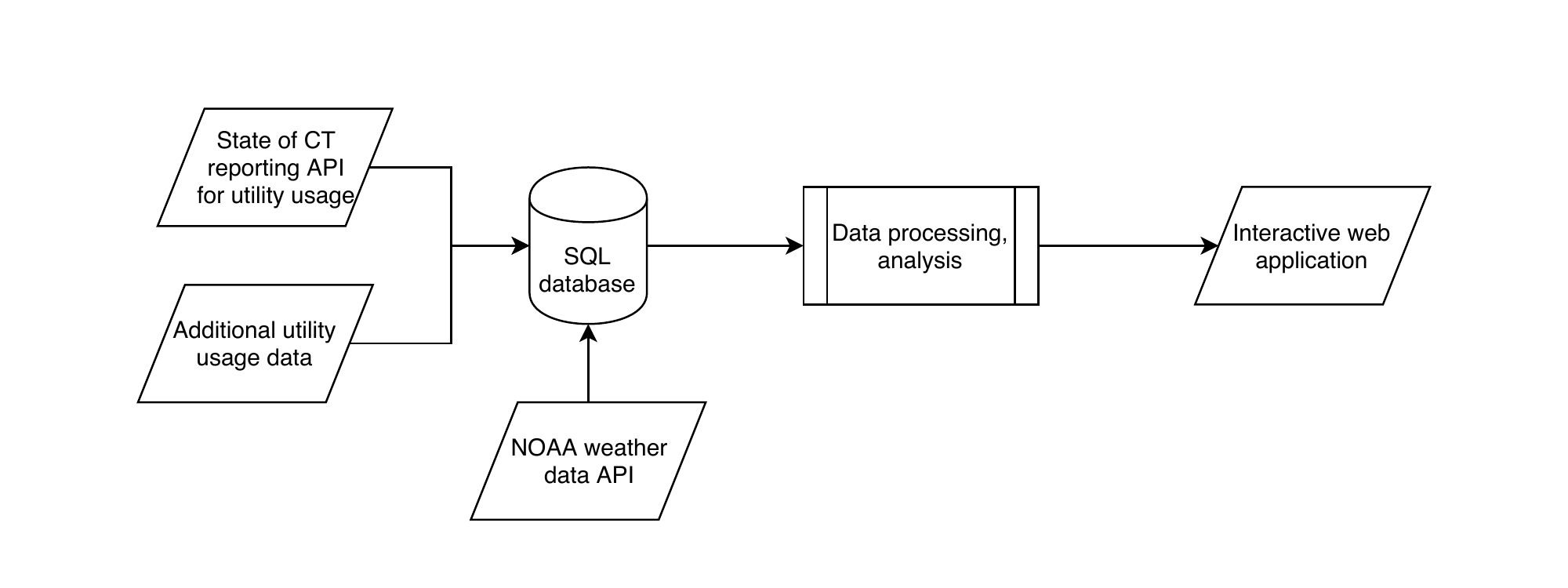}
  \caption{Flow diagram of data ingestion, storage, processing, and
    presentation to operators.}
  \label{fig:implementation}
\end{figure}

Data management and analysis proceed from records of monthly utility
usage, in terms of the specific dataset discussed above, as well as a
suite of additional datasets collected on separate utility services
including electric, water, and sewage.  We implemented an automated
software procedure to periodically download data from a state-wide
reporting system used to track state utility records.  After
extracting utility data via a web API, we store inputs and outputs in
a database server.  We also integrated into the same database system
those datasets not yet available through the web-based procedure.  The
procedures were hosted on a Linux server and backed by a Microsoft SQL
Server database.  We implemented the automation procedures in a
combination of shell scripts, Python, and R. These procedures included
automatic merging of the utility data with NOAA weather data,
normalization and degree day-adjustment, and the analysis detailed in
the previous sections.  A basic flow diagram of the various inputs is
shown in Figure~\ref{fig:implementation}.

\begin{figure}[H]
  \centering
  \fbox{\includegraphics[width=\textwidth]{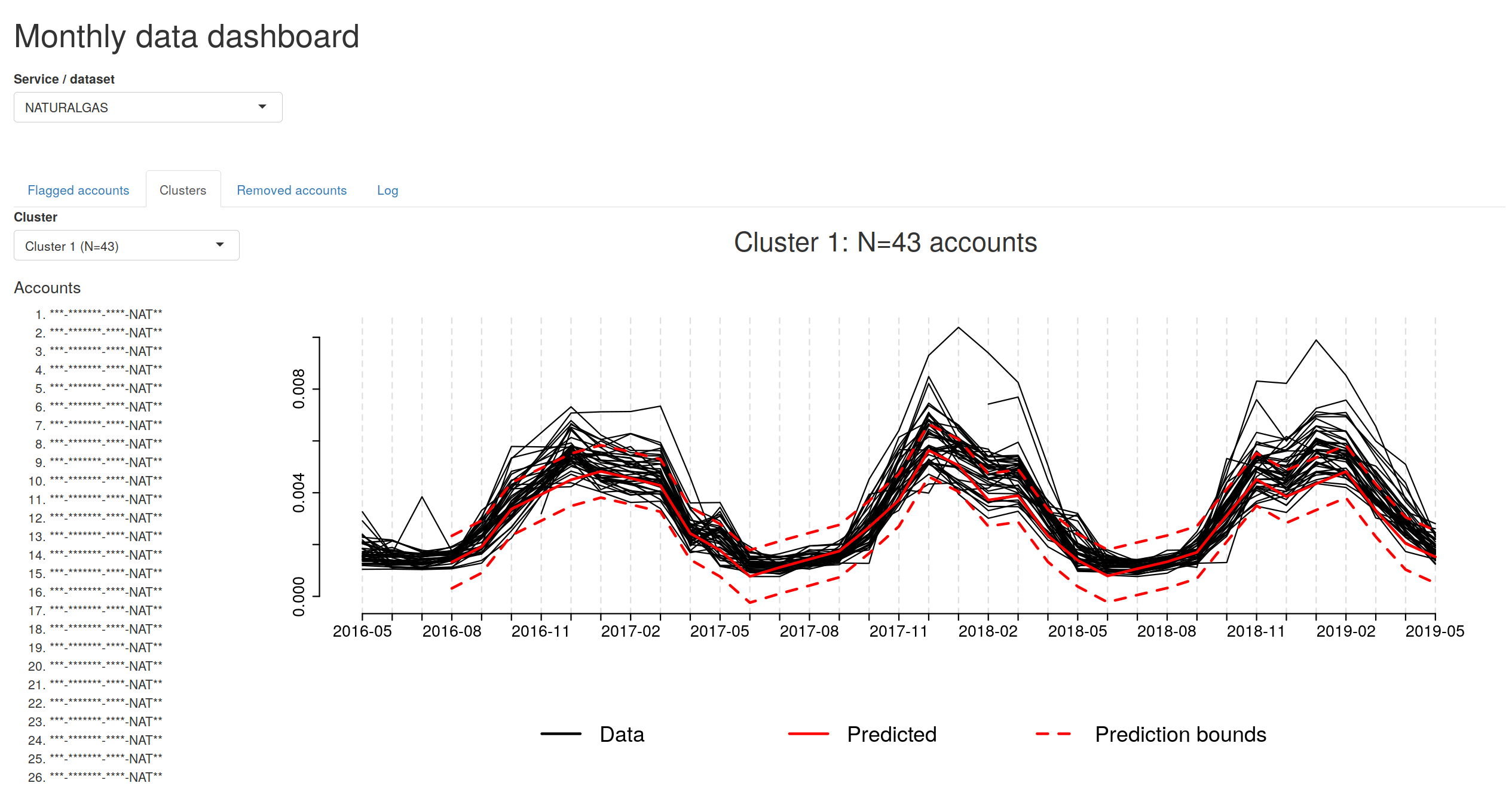}}
  \caption{Cluster analysis output for flagged accounts. Account names
    and numbers are redacted for privacy purposes.}
  \label{fig:flag}
\end{figure}

Following the pre-processing and analysis steps, we provided a web
interface for dynamic interaction with the data and statistical
outputs.  These primarily consisted of a database with a record of
which utility accounts were flagged as anomalous, which Facilities Operations
engineers interact with through a dashboard interface implemented in R
Shiny \citep{shiny}. This software implementation provides
professional-quality, contemporary web technology, as well as back-end
implementation that may be managed, customized, and extended by
statisticians who were not previously trained in web development.  We
implemented the flagging system in such a way that Facilities Operations
may dismiss flags after review of the physical facilities, which was
recorded in the database.  The cluster results of the flagging
interface are shown in Figure~\ref{fig:flag}.

Aside from interaction with the flagging system, the primary mechanism
for operator feedback is through the composition of the homogeneous
groups to which the boxplots based analyses are applied.  Although the cluster analysis
provides a good option in the absence of operator-derived groupings,
it is improved by operation in tandem with the extensive domain
experience of the Facilities Operations engineers. Therefore, Statistics and
Facilities Operations can work in tandem to review and verify the validity of the
groupings, as well as the sensibility of the resultant analysis. In
some instances, such as the laundry dryer in Apartment Complex A,
Building 3, it is necessary to manually adjust even the known
homogeneous groups to more accurately reflect the ground truth of
similarity in energy consumption profiles of accounts in the same
group.

\section{Discussion} \label{sec:discussion}

We have described the development and  implementation of easy to understand statistical methods in an applied
context.
The tools and approaches described in this article offer a %robust 
useful framework for analyzing and monitoring energy usage on a university campus through 
an ongoing collaboration between members of the Statistics Department and 
%considering the university's operational and facilities data,
%collaborate with the 
Facilities Operations staff.
% and provide easy
%access to the conclusions and underlying data that drive the
%statistical decisions we make.  
The energy monitoring   % in turn, has helped Facilities Operations 
has enabled us to gain deeper insight into energy usage   %on campus
at specific sites of interest and % This new information allows for quick
identify potential problems without requiring extensive manual searches.  
An attractive characteristic of this project is  that an academic
department in an R1 institution has coordinated a long-term problem
solving collaboration with the university's administrative units.

We have proposed both a model-free graphical procedure for detecting anomalous values for monthly energy usage as well as for monthly proportions within a given year. 
We have also formulated a simple model-based approach in the Bayesian framework which is useful even with small sample sizes. 
%accommodate small sample sizesand shown how this comapres with the model-free method. 
These approaches have been automated and can also be  easily modified for similar 
operational problems that can benefit from applying simple  and informative statistical methods
to a large number of observational units. Our project showcases  well-informed statistical practice in
collaboration with non-technical collaborative counterparts.

 %Working together with non-statisticians, we have identified suitable transformations of raw data to permit anomaly detection on
%yearly proportions. 
%We did so in a model-free fashion, by
%using boxplots of groups of similar accounts to establish outlier
%limits. We proposed a basic linear model for the logit-transformed
%values of the data, which we formulated in the Bayesian framework to
%accommodate small sample sizes.  Compared with the model-free
%procedure, the model-based approach required substantially more
%detailed specification of proposed outliers to detect, without
%usefully capturing the group structures that persist across
%accounts. The model-free approach, in contrast, offers greater power
%to identify a wide range of possible outlier patterns, while using
%simple graphical methods.  This reduced the theoretical machinery
%needed to understand, communicate, and use the procedure. In this way,

\section*{Acknowledgements}

The authors would like to thank the Utility Operations \& Energy
Management team at Facilities Operations, especially Mark Bolduc and
Brian McKeon for their time, for their time, resources, and expertise.
We would like to thank the Editor, the Associate Editor, and two
reviewers for their helpful comments and suggestions, which led to an
improved version of the paper.

\bibliographystyle{apalike}
%\bibliography{ref.bib}
\bibliography{ref}

\end{document}